\documentclass{article}

\usepackage{arxiv}

\usepackage[utf8]{inputenc} 
\usepackage[T1]{fontenc}    
\usepackage{hyperref}       
\usepackage{url}            
\usepackage{booktabs}       
\usepackage{amsfonts}       
\usepackage{nicefrac}       
\usepackage{microtype}      
\usepackage{lipsum}
\usepackage{graphicx}
\graphicspath{ {./images/} }

\usepackage{comment}
\usepackage{hyperref}

\title{Predicting Sleeping Quality using Convolutional Neural Networks}

\author{
 Vidya Rohini Konanur Sathish\\
  Department of Computer and Information Sciences\\
  Northumbria University\\
  Newcastle upon Tyne, NE1 8ST \\
  \texttt{vidya.sathish@northumbria.ac.uk} \\
   \And
 Wai Lok Woo\\
  Department of Computer and Information Sciences\\
  Northumbria University\\
  Newcastle upon Tyne, NE1 8ST \\
  \texttt{wailok.woo@northumbria.ac.uk} \\
  \And
 Edmond S. L. Ho\thanks{corresponding author}\\
  Department of Computer and Information Sciences\\
  Northumbria University\\
  Newcastle upon Tyne, NE1 8ST \\
  \texttt{e.ho@northumbria.ac.uk} \\
}

\begin{document}
\maketitle
\begin{abstract}
Identifying sleep stages and patterns is an essential part of diagnosing and treating sleep disorders. With the advancement of smart technologies, sensor data related to sleeping patterns can be captured easily. In this paper, we propose a Convolution Neural Network (CNN) architecture that improves the classification performance. In particular, we benchmark the classification performance from different methods, including traditional machine learning methods such as Logistic Regression (LR), Decision Trees (DT), k-Nearest Neighbour (k-NN), Na\"ive Bayes (NB) and Support Vector Machine (SVM), on 3 publicly available sleep datasets. The accuracy, sensitivity, specificity, precision, recall, and F-score are reported and will serve as a baseline to simulate the research in this direction in the future.
\end{abstract}


\section{Introduction}
\label{sec:introduction}


The collection of data by numerous healthcare sensors assisted the Artificial Intelligence (AI) systems in predicting and analysing various types of health-related issues \cite{Lam:2019}. 
Such input data can be used as input for Machine Learning (ML) algorithms to further analyze the patients' health status automatically \cite{Ho:HealthcareIoT2022}. Deep Learning (DL) is a trending expansion of the classical neural network method where a greater number of complex non-linear data patterns can be explored. Another side of the popularity of DL is that complex operations and computations on data can be executed easily \cite{Jiang230}. Deep learning algorithms like multi-layer perceptron, Recurrent Neural Network (RNN) and Convolution Neural Network (CNN) are applied successfully in various domains to solve challenging tasks. The elementary calculation unit in neural networks is a perceptron that achieves a linear combination of input features accompanied by a nonlinear transformation \cite{Biswal:2018}.

Recently, the understanding of health and well-being has improved in the society including quality of sleep, eating habits and physical activities. It is important to understand the relationship between health and sleep quality as they are heavily linked to each other. Sleep is an essential physiological activity of the human body and sleep quality will affect significantly our health \cite{ZHANG2018181}. The reduced sleep causes numerous sleep disorders such as insomnia, narcolepsy, sleep apnea, etc., affecting the overall health \cite{Anishchenko:2019}. Sleep disorder can be diagnosed using Polysomnography (PSG) which records the Electroencephalography (EEG) signals at various locations over the head, electromyography (EMG), electrooculography (EOG) signals and many more. There are numerous times series data that are recorded over the night and every 30-second time segment will be allocated to a human sleep expert for a sleep stage analysis using reference nomenclature like those suggested by the American Academy of Sleep Medicine (AASM) \cite{Chambon:2018}. Sleep apnea is a decreased or complete disturbance of breathing for a minimum of 10 seconds and it is a regularly observed issue among sleep disorders. Sleep apnea can be categorised into three types: obstructive, central and mixed. The breathing interruption of airflow causes the human body not to generate the basic necessary hormones and affects the life of an individual in unrestful, unhealthy and unbearable conditions \cite{Timus:2017}.

An evaluation of sleep quality is therefore needed so that sleep disorders can be detected at the initial stages. The reason behind poor sleep quality is an accomplice with anxiety, physical activity, financial pressure, working hours, smoking, alcohol consumption, and stress. Numerous researchers have predicted that the changes in physical activity are connected to the changes in the rigidity of sleep disorder by causing breathing difficulties followed by disturbed or poor sleep \cite{Perez-Pozuelo2020}. The diagnosis and treatment of sleep-related diseases should be effective and prominently rely on the accurate classification of sleep stages. Therefore, sleep stage classification plays an important role in sleep analysis. Some of the existing work focuses on measuring how young adults convey their sleep habits on social media, as well as how the social media lifestyle is linked to the quality of sleep. Garett et al. \cite{Garett2018} have associated the usage of electronic media and the sleep time and quality that has affected young adults.    

The accessibility and scope of digital technologies for sleep measurement have drastically grown in recent years. Both medical and consumer-grade smart devices (remote sensing, mobile health, wearable gadgets-fitness tracker) across a range of areas are becoming more advanced and affordable. After the sleep data is pre-processed, the data modelling will be initiated for further analyzing the data. With the advancement of Deep Learning in the last decade, the implementation of Artificial Neural Network (ANN) has had a positive impact on the health-related industry. A well-recognised algorithm among several deep learning models is CNN which is a leading technique in computer vision tasks \cite{Yamashita2018}.

In this paper, we propose a Convolution Neural Network (CNN) architecture that improves the classification performance. In particular, we benchmark the classification performance from different methods, including traditional machine learning methods such as Decision Trees (DT), k-Nearest Neighbour (k-NN), Na\"ive Bayes (NB) and Support Vector Machine (SVM), on 3 publicly available sleep datasets. The accuracy, sensitivity, specificity, precision, recall, and F-score are reported and will serve as a baseline to simulate the research in this direction in the future.

The contributions of this research can be summarized as:
\begin{itemize}
    \item We propose 2 new CNN architectures for predicting sleeping quality from a wide range of data captured from smart sensors and surveys.
    \item We conducted experiments to evaluate the performance of the proposed CNNs and compared them with traditional machine learning algorithms on 3 public datasets.
\end{itemize}

\section{Related Work}

Mental stress is one of the major problems that lead to many other diseases, such as sleep disorders \cite{Machado:2018}. Analyzing sleeping patterns and the reasons that are leading to sleep disorders or insomnia is an active research area. For example, Garett et al. \cite{Garett2018} analyze the relationship between the trending technologies (such as social media) and the quality of sleep. The sleep duration and quality are also dependent upon the physical activity performed that affects the production of the hormones in the body 
\cite{Seixas2018}. 

The identification of sleep stages plays a vital role in discovering sleep quality, which is estimated by analyzing the polysomnography (PSG) reports. A PSG study performed in laboratories consists of the analysis of electroencephalogram (EEG), electromyogram (EMG), and electrooculogram (EOG). 
These physiological signals are calculated by sensors that are attached to the patient’s body \cite{Dafna2018}. As per the American Academy Of Sleep Medicine (AASM) rules, five different sleep stages are identified – Wake (W), Rapid Eye Movements (REM), Non-REM1 (N1), Non-REM2 (N2) and Non-REM (N3) as well as slow-wave sleep or even deep sleep. 

Traditional sleep scoring algorithms either from actigraphy signals or PSG are likely to be created from an experimental perspective. These heuristic methods are built on prior experience/knowledge of sleep physiology and detecting modality \cite{Perez-Pozuelo2020}. Whereas, the estimation of traditional and ML algorithms is presented by applying the standard quality metrics like accuracy, recall, and precision for each dataset.  By improving clinical metrics, ML techniques facilitate the doctors to be better informed and will be able to make appropriate clinical decisions \cite{Perez-Pozuelo2020}.

There are wide range of models and methods that are suggested for sleep pattern recognition with self-supervised learning, for example, Zhao et al. \cite{Zhao:2020} proposed a framework consisting of the recognition tasks including upstream and downstream modules. The upstream task is composed of pre-training and feature depiction phases. The extraction of frequency-domain and rotation feature sets to develop new labels simultaneously with the original data. The downstream task is a dynamic Bidirectional Long-Short Term Memory (BiLSTM) module for modelling the transient sleep data \cite{Zhao:2020}. 
In the recent decade, unobtrusive sleep monitoring is one of the popular topics for sleep pattern recognition. Most of the surveys are interested in the sleep stage classification that is dependent on the study of cardiorespiratory features 
\cite{Tataraidze:2017}.
  
A recent study has shown that it is possible to estimate the sleep structure based on the respiratory parameters or interpretation of heart rate variability. The sleep stage classification is based on the respiratory pattern, as it is recorded and dependent upon the body surface displacement caused by respiratory motions. The body surface displacement is recorded in a non-contact approach by Bioradiolocation (BRL) which is a remote detection method of limb and organ motions by using a radar \cite{Tataraidze:2018}.  

Obstructive Sleep Apnea (OSA) is the prevalent kind of sleep breathing disorder and it is represented by repetitive incidents of partial or complete barriers or obstructions while sleeping, generally linked with a decreased blood oxygen saturation. 
Rodrigues et al. \cite{Rodrigues:2020} proposed to focus on the data pre-processing approach with an exhaustive feature selection, and evaluated the method using over 60 regression and classification algorithms in the experiments. 
Predictive models are needed to assist the clinicians in the diagnosis of OSA using Home Sleep Apnea Tests (HSATs). While there are two types of etiologies of non-diagnostic HSATs, both require a referral for PSG. In \cite{Stretch2019}, machine learning technique is used for predicting non-diagnostic HSATs. Compared to traditional statistical models like logistic regression, machine learning algorithms have stronger predictive power whereas by incurring a decrement in the capability to draw interpretations regarding the relationships between the variables \cite{Stretch2019}.


For classifying sleep stages, various approaches have been proposed in the literature. For example, features can be obtained using the Time-Frequency Image (TFI) representation of EEG signals and the sleep stages are then classified by a Multi-Class Least Squares Support Vector Machine (MC-LSSVM) classifier. The statistical measures of EEG epochs are recorded onto a dense network for feature extraction and a k-means classifier is applied for sleep stage classification. By using a phase encoding algorithm, the complex-valued non-linear features are obtained and used as the input to Complex-Valued Neural Network (CVNN) for classifying sleep stages. A feature extraction method identified as the statistical behaviour of local extrema is indicated for the sleep stage classification. The statistical features calculated from the segmented EEG-epochs are plotted onto the graph, then the structural comparison properties of the graph are classified based on the sleep stages with k-means clustering \cite{TARAN2020105367}.

\section{Datasets}
To evaluate the classification performance of the different machine and deep learning methods, 3 publicly available sleep datasets are downloaded from Kaggle ( \url{https://www.kaggle.com/}). The details of the datasets are explained in the rest of this section. 


\subsection{Dataset 1: Sleep-Study}
This is a survey-based study of the sleeping habits of individuals within the United States of America \cite{Study-Sleep}. The dataset consists of 104 sleeping habits records of individuals. Each record contains six attributes, namely {\it Enough (yes/no), Hours, Phone Reach (yes/no), Phone Time (yes/no), Tired (1 being not tired, 5 being very tired), and Breakfast (yes/no)}. The task is to predict whether the participant has {\it enough} sleep based on the rest of the attributes (i.e. 5) as input features.


\subsection{Dataset 2: Sleep Deprivation}
This dataset \cite{Sleep-Deprivation} is composed of 86 sleep records and each record has 80 attributes that are related to the demographic background of the participant and the response to the Karolinska Sleep Questionnaire. Only 7 key variables are selected as the input feature set, including {\it Age group, Anxiety rate, Depression rate, Panic, Worry, Health problems, and Nap duration}, to predict the {\it Overall sleep quality} and whether the participant has {\it enough} sleep.



\subsection{Dataset 3: Sleep Cycle Data}
This dataset \cite{Sleep-Cycle-Data} consists of 50 sleep records with 8 attributes, namely {\it Start, End, Sleep Quality, Time in Bed, Wake up, Sleep notes, Heart rate, and Activity (Number of Steps per day)}. Since the {\it Sleep note} is a text-based field for the participant to provide a textual description of the activities on that day, this attribute is not included in the classification and the rest of the attributes are used as input features to predict the {\it Sleep Quality}.

\subsection{Data pre-processing}
After the data collection process, each sleep dataset is divided into training and testing sets on a 50-50 data split. We further performed data normalization on the data to facilitate the classifier training process. 

\section{Methodology}
\begin{figure*}[!h]
  \vspace{-0.2cm}
  \centering
   \includegraphics[width=0.8\linewidth]{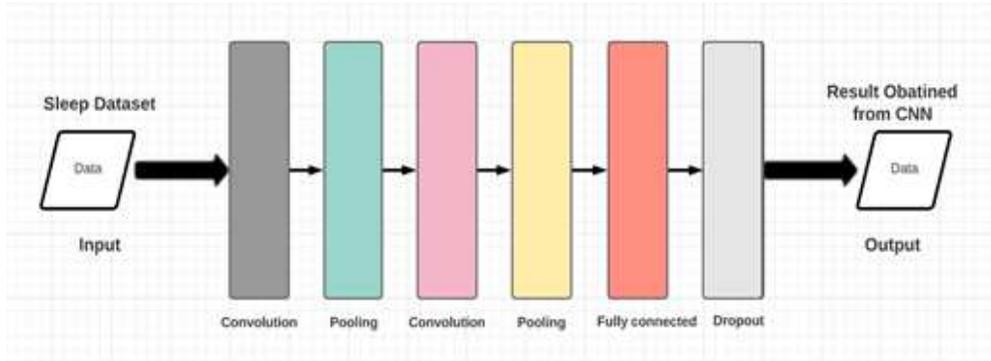}
  \caption{The general network architecture of the proposed CNNs.}
  \label{fig:CNN}
  \vspace{-0.1cm}
\end{figure*}

In this section, the proposed CNN architectures will be introduced. Inspired by the encouraging results in classifying infant movements using 1D and 2D CNNs \cite{McCay:DeepBaby}, we propose a general CNN framework for classifying the sleep data and the network architecture is illustrated in \ref{fig:CNN}. 

Specifically, the initial input layer is followed by a convolution layer and then the ReLU activation layer. A Max Pooling layer is then added for obtaining a more abstract representation from the input. Such a Conv-ReLU-MaxPool structure is repeated before feeding the deep representation into a Fully-connected layer which is then followed by a dropout layer before the classification. Finally, the output layer comprises of softmax layer and the predicted class label can be obtained.


Overfitting is a known challenge when training machine learning models with small datasets such as those being used in this research. At the initial stage of experimenting the training, validation and testing performance were very poor. By adding dropout layers, the impact of overfitting was alleviated.

\subsection{1D Convolutional Neural Networks}
Here, we propose 2 1D CNN architectures based on the general structure illustrated in Figure \ref{fig:CNN}. The 2 new networks, namely $CONV-1D_1$ and $CONV-1D_2$, share the same network architecture but with different dimensional in the intermediate layers. Specifically, the Max Pooling layers in $CONV-1D_1$ reduce the dimensionality of the intermediate representation while $CONV-1D_2$ will preserve the size of the representation as in the output of the convolutional layer. Such a design enables us to evaluate how the Max Pooling affects the performance of the CNNs. 


\section{Experimental Results}

\subsection{Evaluation Metrics} \label{sec:metrics}
Several metrics are used for evaluating the classification performance in this research, including accuracy (AC), sensitivity (SE) and specificity (SP) which are calculated from true positive (TP), true negative (TN), false positive (FP) and false negative (FN). We also calculate the precision (PR), recall (RE), and F1 Score (F1). The equations for the calculation are as follows:
\begin{equation}
	SE = {\frac{TP}{TP+FN}} 
\end{equation}

\begin{equation}
	SP = {\frac{TN}{TN+FP}} 
\end{equation}

\begin{equation}
	AC = {\frac{TP+TN}{TP+FN+TN+FP}} 
\end{equation}

\begin{equation}
	PR = {\frac{TP}{TP+FP}} 
\end{equation}

\begin{equation}
	RE = {\frac{TP}{TP+FN}} 
\end{equation}

\begin{equation}
	F1 = {2\cdot\frac{PR \cdot RE}{PR + RE}} 
\end{equation}

\subsection{Classification Results}
In the first experiment, we report the performance of the Logistic Regression Classifier on the 3 datasets as an overview. Several evaluation metrics were used as stated in Section \ref{sec:metrics}. The results are presented in Table \ref{tab:3_datasets}. It can be seen that the datasets are challenging in general, with a classification accuracy ranging from 55.11\% to 63.46\%. With the F1 scores, the best performance (59.01\%) was obtained on the Sleep-Study dataset which is the largest one among the 3 datasets. In contrast, the lowest performance (51.32\%) was obtained in the Sleep Deprivation dataset.

\begin{table*}[htb]
\caption{The performance of the logistic regression classifier on the three datasets.}\label{tab:3_datasets} \centering
\begin{tabular}{c c c c}
  \hline
  & Sleep-Study & Sleep Deprivation & Sleep Cycle Data \\
  \hline
  AC & 63.46\% & 58.02\% & 55.11\% \\
  SE & 64.52\% & 62.41\% & 57.02\% \\
  SP & 61.90\% & 59.85\% & 56.45\% \\
  F1 & 59.01\% & 51.32\% & 53.33\% \\
  PR & 66.67\% & 56.13\% & 52.23\% \\
  RE & 52.97\% & 65.44\% & 49.36\% \\
  \hline
\end{tabular}
\end{table*}

In the second experiment, we further evaluate the classification performance using different classifiers \cite{McCay:EMBC2019,McCay:TNSRE2022}. Table \ref{tab:ml_results} reports the accuracy obtained using different methods on the 3 datasets for further evaluating the classification performance. For decision trees, 56.22\%, 61.26\% and 51.33\% were obtained from the 3 datasets. For k-NN classifier with k=1 is being used, the accuracies were 55.33\%, 59.50\% and 55.94\%, respectively. When the k=10 is used with k-NN, 61.83\%, 64.70\% and 58.11\% were obtained. It can be seen that performance was improved when k=10 is used instead of k=1. In addition, Na\"ive Bayes achieved 59.65\%, 53.69\% and 56.03\% on the datasets. 
The SVM classifier achieved 59.21\%, 62.66\% and 59.26\% on the three datasets respectively. Compared to decision trees, k-NN and Gaussian Na\"ive Bayes classification models of machine learning, the SVM classifier performed better on all three sleep datasets.

For the proposed CNN-based classifiers, $CONV-1D_1$ obtained the accuracies of 59.18\%, 64.54\% and 59.00\% on the 3 datasets. $CONV-1D_2$ performed slightly better than $CONV-1D_1$ with 61.22\%, 65.23\% and 58.73\% obtained from the datasets, which shows the intermediate representations have to be in higher dimensionality in order to better model the data. It can be seen that $CONV-1D_1$ and $CONV-1D_2$ have shown better performances, more robust and more consistent than the traditional machine learning approaches. Although Logical Regression and SVM performed the best on Sleep-Study and Sleep Cycle Data datasets, the proposed CNN-based methods are more consistent on all 3 datasets and obtained comparable accuracy with those two methods.


\begin{table*}[htb]
\caption{The performance of different classifiers on the three datasets.}\label{tab:ml_results} \centering
\begin{tabular}{c c c c}
  \hline
  & Sleep-Study & Sleep Deprivation & Sleep Cycle Data \\
  \hline
  Logical Regression & {\bf 63.46\%} & 58.02\% & 55.11\% \\
  Decision Tree & 56.22\% & 61.26\% & 51.33\% \\
  k-NN (k=1) & 55.33\% & 59.40\% & 55.94\% \\
  k-NN (k=10) & 61.83\% & 64.70\% & 58.11\% \\
  Na\"ive  Bayes & 53.69\% & 59.65\% & 56.03\% \\
  SVM & 59.21\% & 62.66\% & {\bf 59.26\%} \\
  \hline
  $CONV-1D_1$ & 59.19\% & 64.54\% & 59.00\% \\
  $CONV-1D_2$ & 61.22\% & {\bf 65.23\%} & 58.73\% \\
  \hline
\end{tabular}
\end{table*}

\section{Conclusions}
\label{sec:conclusion}
In this paper, we proposed new CNN classification frameworks for predicting sleeping quality from a wide range of data captured from smart sensors and surveys. To show the effectiveness of the proposed CNN architectures, we evaluate the proposed classification framework on 3 publicly available datasets. It is a challenging task since the datasets are small in general. We further conducted the experiments using traditional classifiers including Logistic Regression, Decision Trees, k-Nearest Neighbour, Na\"ive Bayes and Support Vector Machine as baselines for comparison. Experimental results highlighted the robustness of the proposed CNN architectures with highly consistent results obtained from different datasets.

\bibliographystyle{unsrt}  
\bibliography{references}  

\begin{thebibliography}{10}

\bibitem{Lam:2019}
Pham~Son Lam, Nguyen Dinh~Son, Hoang~Phuong Chi, Nguyen~Thi Phuoc~Van, and
  Nguyen Duc~Minh.
\newblock Novel algorithm to classify sleep stages.
\newblock In {\em 2019 13th International Conference on Sensing Technology
  (ICST)}, pages 1--6, 2019.

\bibitem{Ho:HealthcareIoT2022}
Edmond S.~L. Ho.
\newblock {\em Data Security Challenges in Deep Neural Network for Healthcare
  IoT Systems}, pages 19--37.
\newblock Springer International Publishing, Cham, 2022.

\bibitem{Jiang230}
Fei Jiang, Yong Jiang, Hui Zhi, Yi~Dong, Hao Li, Sufeng Ma, Yilong Wang, Qiang
  Dong, Haipeng Shen, and Yongjun Wang.
\newblock Artificial intelligence in healthcare: past, present and future.
\newblock {\em Stroke and Vascular Neurology}, 2(4):230--243, 2017.

\bibitem{Biswal:2018}
Siddharth Biswal, Haoqi Sun, Balaji Goparaju, M.~{Brandon Westover}, Jimeng
  Sun, and {Matt T.} Bianchi.
\newblock Expert-level sleep scoring with deep neural networks.
\newblock {\em Journal of the American Medical Informatics Association},
  25(12):1643--1650, January 2018.

\bibitem{ZHANG2018181}
Junming Zhang and Yan Wu.
\newblock Complex-valued unsupervised convolutional neural networks for sleep
  stage classification.
\newblock {\em Computer Methods and Programs in Biomedicine}, 164:181--191,
  2018.

\bibitem{Anishchenko:2019}
Lesya Anishchenko, Andrey Zhuravlev, Vladimir Razevig, Margarita Chizh, Kseniya
  Evteeva, Lyudmila Korostovtseva, Mikhail Bochkarev, and Yurii Sviryaev.
\newblock Non-contact sleep disorders detection framework for smart home.
\newblock In {\em 2019 PhotonIcs Electromagnetics Research Symposium - Spring
  (PIERS-Spring)}, pages 3553--3557, 2019.

\bibitem{Chambon:2018}
Stanislas Chambon, Mathieu~N. Galtier, Pierrick~J. Arnal, Gilles Wainrib, and
  Alexandre Gramfort.
\newblock A deep learning architecture for temporal sleep stage classification
  using multivariate and multimodal time series.
\newblock {\em IEEE Transactions on Neural Systems and Rehabilitation
  Engineering}, 26(4):758--769, 2018.

\bibitem{Timus:2017}
Oğuz~Han TİMUŞ and Emine~Doğru BOLAT.
\newblock K-nn-based classification of sleep apnea types using ecg.
\newblock {\em Turkish Journal of Electrical Engineering and Computer Science},
  25(4):3008--3023, 2017.

\bibitem{Perez-Pozuelo2020}
Ignacio Perez-Pozuelo, Bing Zhai, Joao Palotti, Raghvendra Mall, Micha{\"e}l
  Aupetit, Juan~M. Garcia-Gomez, Shahrad Taheri, Yu~Guan, and Luis
  Fernandez-Luque.
\newblock The future of sleep health: a data-driven revolution in sleep science
  and medicine.
\newblock {\em npj Digital Medicine}, 3(1):42, Mar 2020.

\bibitem{Garett2018}
Renee Garett, Sam Liu, and Sean~D. Young.
\newblock The relationship between social media use and sleep quality among
  undergraduate students.
\newblock {\em Information, communication and society}, 21(2):163--173, 2018.

\bibitem{Yamashita2018}
Rikiya Yamashita, Mizuho Nishio, Richard Kinh~Gian Do, and Kaori Togashi.
\newblock Convolutional neural networks: an overview and application in
  radiology.
\newblock {\em Insights into Imaging}, 9(4):611--629, Aug 2018.

\bibitem{Machado:2018}
José~Raúl {Machado Fernández} and Lesya Anishchenko.
\newblock Mental stress detection using bioradar respiratory signals.
\newblock {\em Biomedical Signal Processing and Control}, 43:244--249, 2018.

\bibitem{Seixas2018}
Azizi~A. Seixas, Dwayne~A. Henclewood, Stephen~K. Williams, Ram Jagannathan,
  Alberto Ramos, Ferdinand Zizi, and Girardin Jean-Louis.
\newblock Sleep duration and physical activity profiles associated with
  self-reported stroke in the united states: Application of bayesian belief
  network modeling techniques.
\newblock {\em Frontiers in neurology}, 9:534--534, Jul 2018.

\bibitem{Dafna2018}
Eliran Dafna, Ariel Tarasiuk, and Yaniv Zigel.
\newblock Sleep staging using nocturnal sound analysis.
\newblock {\em Scientific Reports}, 8(1):13474, Sep 2018.

\bibitem{Zhao:2020}
Aite Zhao, Junyu Dong, and Huiyu Zhou.
\newblock Self-supervised learning from multi-sensor data for sleep
  recognition.
\newblock {\em IEEE Access}, 8:93907--93921, 2020.

\bibitem{Tataraidze:2017}
Alexander Tataraidze, Lesya Anishchenko, Lyudmila Korostovtseva, Mikhail
  Bochkarev, Yurii Sviryaev, and Sergey Ivashov.
\newblock Estimation of a priori probabilities of sleep stages: A cycle-based
  approach.
\newblock In {\em 2017 39th Annual International Conference of the IEEE
  Engineering in Medicine and Biology Society (EMBC)}, pages 3745--3748, 2017.

\bibitem{Tataraidze:2018}
Alexander~B Tataraidze, Lesya~N. Anishchenko, Lyudmila~S. Korostovtseva,
  Mikhail~V. Bochkarev, and Yurii~V. Sviryaev.
\newblock Non-contact respiratory monitoring of subjects with sleep-disordered
  breathing.
\newblock In {\em 2018 IEEE International Conference "Quality Management,
  Transport and Information Security, Information Technologies" (IT QM IS)},
  pages 736--738, 2018.

\bibitem{Rodrigues:2020}
Jose~F. Rodrigues, Jean-Louis Pepin, Lorraine Goeuriot, and Sihem Amer-Yahia.
\newblock {\em An Extensive Investigation of Machine Learning Techniques for
  Sleep Apnea Screening}, page 2709–2716.
\newblock Association for Computing Machinery, New York, NY, USA, 2020.

\bibitem{Stretch2019}
Robert Stretch, Armand Ryden, Constance~H. Fung, Joanne Martires, Stephen Liu,
  Vidhya Balasubramanian, Babak Saedi, Dennis Hwang, Jennifer~L. Martin,
  Nicol{\'a}s Della~Penna, and Michelle~R. Zeidler.
\newblock Predicting nondiagnostic home sleep apnea tests using machine
  learning.
\newblock {\em Journal of clinical sleep medicine : JCSM : official publication
  of the American Academy of Sleep Medicine}, 15(11):1599--1608, Nov 2019.

\bibitem{TARAN2020105367}
Sachin Taran, Prakash~Chandra Sharma, and Varun Bajaj.
\newblock Automatic sleep stages classification using optimize flexible
  analytic wavelet transform.
\newblock {\em Knowledge-Based Systems}, 192:105367, 2020.

\bibitem{Study-Sleep}
Michael Lomuscio.
\newblock Sleep study - a survey-based study of the sleeping habits of
  individuals within the us, 2019.
\newblock https://www.kaggle.com/mlomuscio/sleepstudypilot.

\bibitem{Sleep-Deprivation}
Frederick Feraco.
\newblock Sleep deprivation, 2018.
\newblock https://www.kaggle.com/feraco/sleep-deprivation.

\bibitem{Sleep-Cycle-Data}
Kubilay Işen.
\newblock Sleep cycle data, 2020.
\newblock https://www.kaggle.com/robottesla/sleepcycledata.

\bibitem{McCay:DeepBaby}
Kevin~D. McCay, Edmond S.~L. Ho, Hubert P.~H. Shum, Gerhard Fehringer, Claire
  Marcroft, and Nicholas~D. Embleton.
\newblock Abnormal infant movements classification with deep learning on
  pose-based features.
\newblock {\em IEEE Access}, 8:51582--51592, 2020.

\bibitem{McCay:EMBC2019}
K.~D. {McCay}, E.~S.~L. {Ho}, C.~{Marcroft}, and N.~D. {Embleton}.
\newblock Establishing pose based features using histograms for the detection
  of abnormal infant movements.
\newblock In {\em 2019 41st Annual International Conference of the IEEE
  Engineering in Medicine and Biology Society (EMBC)}, pages 5469--5472, July
  2019.

\bibitem{McCay:TNSRE2022}
Kevin~D. McCay, Pengpeng Hu, Hubert P.~H. Shum, Wai~Lok Woo, Claire Marcroft,
  Nicholas~D. Embleton, Adrian Munteanu, and Edmond S.~L. Ho.
\newblock A pose-based feature fusion and classification framework for the
  early prediction of cerebral palsy in infants.
\newblock {\em IEEE Transactions on Neural Systems and Rehabilitation
  Engineering}, 30:8--19, 2022.

\end{thebibliography}

\end{document}